\documentclass[aps,prl,twocolumn,amsmath,10pt,superscriptaddress,floatfix,nofootinbib]{revtex4-1}

\usepackage{epsfig,amssymb,amsfonts,amsmath,mathtools,bm,color,xcolor,graphicx,braket,adjustbox,esint,upgreek}
\usepackage{multirow}
\usepackage[utf8]{inputenc}

\newcommand{\bef}{\begin{figure}[htb!]\centering}
\newcommand{\eef}{\end{figure}}

\newcommand{\be}{\begin{equation}}
\newcommand{\ee}{\end{equation}}
\newcommand{\bea}{\begin{eqnarray}}
\newcommand{\eea}{\end{eqnarray}}
\newcommand{\beas}{\begin{eqnarray*}}
\newcommand{\eeas}{\end{eqnarray*}}

\newcommand{\x}{\rm{X}(3872)}

\newcommand{\scnu}{\affiliation{Guangdong Provincial Key Laboratory of Nuclear Science,\\ Institute of Quantum Matter, South China Normal University, Guangzhou 510006, China}}

\newcommand{\scnuhk}{\affiliation{Guangdong-Hong Kong Joint Laboratory of Quantum Matter, Southern Nuclear Science Computing Center, South China Normal University, Guangzhou 510006, China}}

\newcommand{\indiana}{\affiliation{Physics Department and Center for Exploration of Energy and Matter,\\ Indiana University, 2401 N Milo B. Sampson Lane, Bloomington, Indiana 47408, USA}}

\newcommand{\ihep}{\affiliation{Theoretical Physics Center for Science Facilities, \\ Institute of High Energy Physics, Chinese Academy of Sciences, Beijing 100049, China}}

\graphicspath{{Figures/}}

\begin{document}

\title{Deciphering the Nature of X(3872) in Heavy Ion Collisions}

\author{Hui Zhang}\email{Mr.zhanghui@m.scnu.edu.cn}
\scnu
\scnuhk

\author{Jinfeng Liao}\email{liaoji@indiana.edu}
\indiana

\author{Enke Wang}\email{wangek@scnu.edu.cn}
\scnu
\scnuhk

\author{Qian Wang}\email{qianwang@m.scnu.edu.cn}
\scnu
\scnuhk
\ihep

\author{Hongxi Xing}\email{hxing@m.scnu.edu.cn}
\scnu
\scnuhk

\begin{abstract}
Exploring the nature of exotic multiquark candidates such as  the $\x$  plays a pivotal role in understanding quantum chromodynamics (QCD). Despite significant efforts, consensus on their internal structures is still lacking.  As a prime example, it remains a pressing open question to decipher the $\x$ state between two popular exotic configurations: a loose hadronic molecule or a compact tetraquark. We demonstrate a novel approach to help address this problem by studying the $\x$ production in heavy ion collisions, where a hot fireball with ample light as well as charm (anti-)quarks is available for producing the exotics.  Adopting a multiphase transport model (AMPT) for describing such collisions and implementing appropriate production mechanism of either molecule or tetraquark picture,  we compute and compare a series of observables  for $\x$ in Pb-Pb collisions at the Large Hadron Collider. We find the fireball volume plays a crucial role, leading to a 2-order-of-magnitude difference in the $\x$ yield and a markedly different centrality dependence between hadronic molecules and compact tetraquarks, thus offering a unique opportunity for distinguishing the two scenarios. We also make the first prediction of $\x$ elliptic flow coefficient to be tested by future experimental measurements. 
\end{abstract}

\maketitle

{\it Introduction.---}
The strong interaction is one of the four basic forces in our Universe, and its underlying theory is known as quantum chromodynamics (QCD). While QCD is based on fundamental particles called quarks and gluons, we can only directly observe hadrons in which quarks or gluons are confined by nonperturbative QCD interactions. To understand the making of all possible hadrons is a core question that has been a persistent challenge to our understanding of QCD~\cite{Shepherd:2016dni,Briceno:2015rlt}. 

The quark model, as a starting point of such inquiry, was known to allow for  multiquark configurations since the very beginning~\cite{Jaffe:1976ig}. 
However, it had been misinterpreted to only contain 
the quark-antiquark mesons and the three-quark baryons for quite a long time, due to
the absent experimental evidence of the hadrons beyond those two configurations. 
The recent observations of the $\x$~\cite{Choi:2003ue} with quantum number $J^{\rm{PC}}=1^{++}$~\cite{Aaij:2015eva},  as the first exotic candidate, 
and other exotic candidates
 afterwards have driven  the whole community to rethink about various possibilities of  ``exotic hadrons'' in QCD.  Comprehensive efforts 
\cite{Eichten:2007qx,Brambilla:2010cs,Esposito:2014rxa,Chen:2016spr,Chen:2016qju,Dong:2017gaw,Esposito:2016noz,Hosaka:2016pey,Olsen:2017bmm,Guo:2017jvc,Kou:2018nap,Cerri:2018ypt,Liu:2019zoy,Brambilla:2019esw,Guo:2019twa} 
have been made  
to predict or measure their existence and properties. 
 However, the nature of these exotic candidates remains an open question with little consensus from the community. 
 Taking the most-studied $\x$ as a prime example, its proximity to the ($D\bar{D}^*+\rm charge ~ conjugate$) threshold indicates its hadronic molecular picture \cite{Tornqvist:2003na}. Besides that, 
 there are also other scenarios, such as 
diquark-antidiquark tetraquark \cite{Jaffe:2004ph,Close:2004ip,Maiani:2004vq}, hybrid \cite{Li:2004sta},
charmonium\cite{Barnes:2003vb,Eichten:2004uh,Suzuki:2005ha}, quantum mixture of $\chi_{c1}(2P)$ and $D^0\bar D^{*0}$ \cite{Meng:2005er}, as well as other configurations , see, e.g., Refs. \cite{Chen:2016spr, Guo:2017jvc, Ali:2017jda, Lebed:2016hpi} for reviews.

While conventionally electron-positron or proton-proton collisions are used to produce and study exotic hadrons, there has been increasing interest   to study such states in heavy ion collisions. Indeed, given the abundant number of quarks and antiquarks for both light and heavy flavors, these collisions appear to provide the ideal environment for exotic hadron production. 
The first study was performed in the coalescence model in comparison with the statistical model \cite{Cho:2010db, Cho:2011ew}. Later on, further improving of the coalescence model \cite{Sun:2017ooe,Sun:2018mqq}, detailed analysis of its transverse momentum distribution \cite{Andronic:2019wva,Cho:2019syk}, the wave function for tetraquark state \cite{Fontoura:2019opw}, and the hadronic effects \cite{Hong:2018mpk} were considered in heavy ion collisions. Possible effect from a hot pion bath at late time~\cite{Cleven:2019cre} and the hadronic effect \cite{Cho:2013rpa} on the properties of the $\x$ were further discussed.  The possible influence of tetraquarks on QCD phase structures was also explored~\cite{Andronic:2017pug, Pisarski:2016ukx}.  More discussions  can be found in recent reviews  \cite{Cho:2017dcy,Braun-Munzinger:2018hat} and references therein.   
 Most recently, the CMS Collaboration reported the first experimental evidence of $\x$ in Pb-Pb collisions at the Large Hadron Collider (LHC) \cite{CMS:2019vma}, making an important first step toward quantitative investigation of exotic hadrons  in heavy ion collisions. While the present study focuses on nucleus-nucleus collisions, the $\x$  production in proton-proton ($pp$) or proton-nucleus collisions ($pA$) could be equally informative. For example, recent LHCb measurements suggest a suppression of $\x$ production in high multiplicity $pp$ collisions and 
a latest theoretical analysis~\cite{Esposito:2020ywk} found this dependence to be in favor of the tetraquark hypothesis.

In this Letter, we explore such an emerging opportunity to study $\x$ production in heavy ion collisions  and report two essential results. We perform a first quantitative computation of $\x$ production  within a realistic bulk evolution model for a series of  heavy ion observables    and make the first prediction of $\x$ elliptic flow, which is critically needed for the ongoing experimental program. This is done by adopting a multiphase transport model (AMPT)~\cite{Lin:2004en} for describing such collisions and implementing   production mechanism of either molecule or tetraquark picture (as illustrated in Fig.~\ref{fig-xhic}). 
Furthermore, our computations suggest a significantly larger yield of the $\x$ as well as a markedly stronger centrality dependence when assuming its nature to be a hadronic molecule as compared with a compact tetraquark. 
This novel finding points at a unique opportunity for deciphering the nature of $\x$ and help address a long-standing hadron physics challenge with heavy ion  measurements, with the predicted difference between the two rival scenarios well beyond current experimental limitation. All these new results 
are readily testable and shall strongly motivate experimental efforts in the near future.

 \bef
\psfig{file=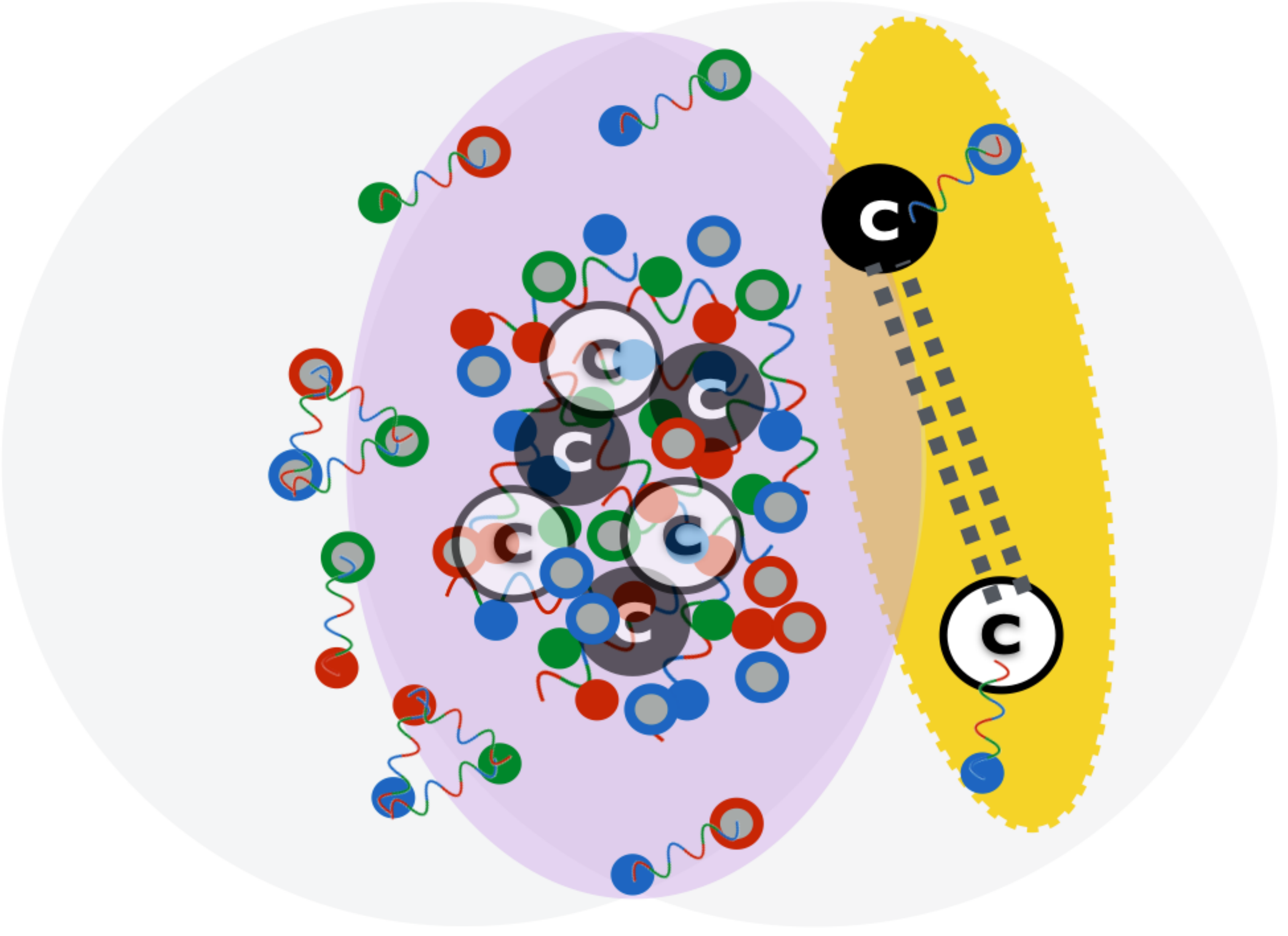, width=1.6in, height=1.5in}\hspace{0.1in}
\psfig{file=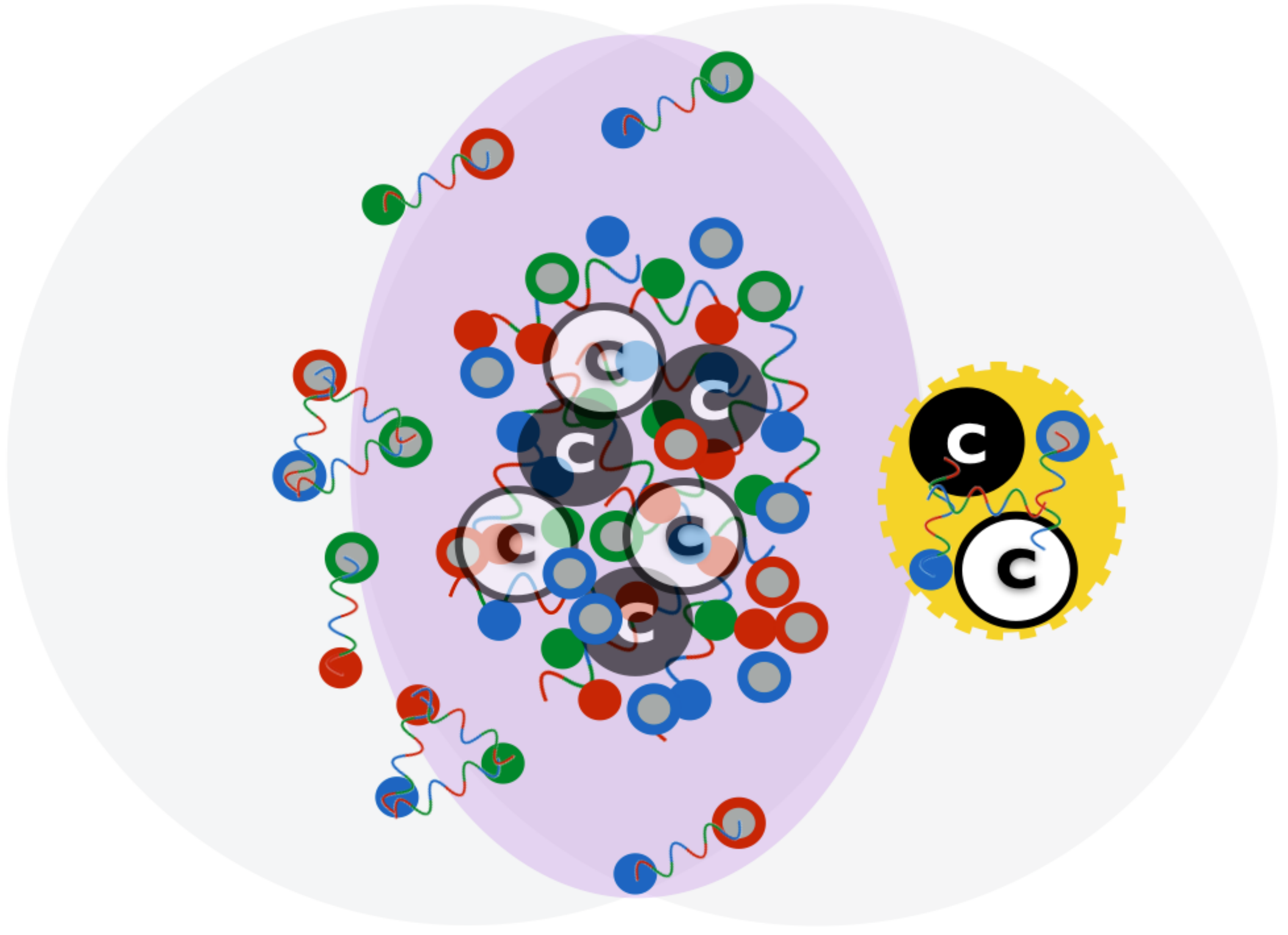, width=1.6in, height=1.5in}
\caption{Illustration of $\x$  production as hadronic molecule (left) or tetraquark  (right) in heavy ion collisions.}
\label{fig-xhic}
\eef

 {\it Framework.---}
In this study, we use the default version of AMPT \cite{Lin:2004en} to estimate the yield of the $\x$ in Pb-Pb collisions at LHC energies. 
AMPT is a widely used event generator to describe the bulk evolution of heavy ion collisions. It incorporates four main components: the fluctuating initial conditions, partonic scatterings modeled by parton cascade, hadronization by using a quark coalescence model, and the subsequent hadronic rescattering. AMPT has been successfully applied to describe a  variety of observables for collision energies ranging from CERN SPS to LHC~\cite{Lin:2014tya, Ma:2016fve, Zhang:2018ucx}. 
In our simulation, we use the settings as in Ref. \cite{Esha:2016svw} which are tuned to describe the elliptic flow for open charm mesons.

The new element we introduce into the AMPT simulations is the mechanism to produce $\x$ for its two possible configurations, i.e., the  hadronic molecular configurations and the   tetraquark configurations. 
Since the $\x$ contains constituent charm quarks or antiquarks, we  need   a reasonable  generation of  individual $c$ and $\bar{c}$ quarks in the partonic phase. This can be calibrated by comparison with experimental data on $D$ meson production in Pb-Pb collisions~\cite{ALICE:2012ab, Adam:2015sza}. It is known that in the default version of AMPT, some of the channels related to initial heavy quark production are missing and efforts to remediate such issues were recently made~\cite{Zheng:2019alz}. We adopt a similar strategy to enhance the initial $c$ and $\bar{c}$ spectra by a factor of $K$, which leads to a reasonable agreement for the total production of 
 $D^+(D^0) + D^{*+}(D^{*0})$ meson between our AMPT results and ALICE measurements for  0-10\% and 30-50\% centralities. 

We next implement the production mechanism for the   hadronic molecule and tetraquark configurations of the $\x$. Both scenarios  stem from  reasonable (albeit drastically different) underlying dynamics~\cite{Bignamini:2009sk,Artoisenet:2009wk,Artoisenet:2010uu,Guo:2014sca,Guo:2017jvc,Albaladejo:2017blx} with supporting evidences and are hard to differentiate at the moment. 
Such a hadronic physics challenge could present an opportunity in heavy ion collisions. Given their rather different structures, one may   expect that their production in heavy ion collisions could   be very different, as  illustrated in Fig.~\ref{fig-xhic}. We consider both possibilities and evaluate $\x$ production in each case accordingly.     
For the   molecule scenario, the $\x$  is formed by the color neutral force, 
as analogy of deuteron first proposed in Ref.~\cite{Tornqvist:2003na},
 between either $D$ and $\bar{D}^*$
or $\bar{D}$ and $D^*$. 
As the effective range expansion works well for a two-body near-threshold system,
  one would expect the $X(3872)$ has an effective range of several fm~[14] from the next-to-leading order contribution.
We use  5-7 {fm} as an illustration. {(Even if we choose the widest plausible range of $2\sim 10$ fm in the molecular picture, the $\x$ yield would only be enhanced by a factor of 3.2 which does not affect the order-of-magnitude estimate.)}
In this case  the ``molecule'' $\x$ is formed in our simulations by coalescence of two  charmed mesons  with   constraints: 5 {fm} $< \rm{relative\ distance} <$ 7 {fm} and $2M_D < \mathrm{pair\ mass} < 2M_{D^*}$.  
For the tetraquark scenario, the $\x$ is formed by a colored force between a color antitriplet diquark $[cq]$ and a color triplet antidiquark $[ \bar{c} \bar{q} ]$~\cite{Maiani:2004vq,Cleven:2015era} analogous to that in a normal meson.
As a result, the tetraquark scenario is of a
  normal hadron size $\lesssim 1\ \mathrm{fm}$.  
 In this case the ``tetra'' $\x$ is formed in our simulations via two steps at freeze-out: (i) First  diquarks ($cq$) and antidiquarks ($\bar{c}\bar q$) are created via partonic coalescence, by matching a $c$ or $\bar{c}$ with the nearest light quark or anti-quark; (ii) Then these diquarks or antidiquarks are further used to form  $\x$ via coalescence by matching the following quantitative constraints: $\rm{relative\ distance} <1 ~\rm{fm}$ and pair mass between
 the upper and lower mass limits of the heavy quark spin partners of the $\x$.
 {(The upper and lower mass limits of the heavy quark spin partners of the $\x$ in tetraquark picture are $4020~\mathrm{MeV}$ and $3780~\mathrm{MeV}$, i.e., the masses of the tetraquarks $|11\rangle_0$ and $|00\rangle_0$ as defined in Refs.~\cite{Maiani:2014aja,Cleven:2015era}, respectively.)}
We note that despite potential differences in the binding energy between molecular and tetraquark pictures, the same mass value should result for the \x  state in both pictures.
 
There is subtlety in forming charmed mesons or (anti-)diquarks with the same flavor contents but different spin composition. In principle one needs to include the spin degrees of freedom to distinguish these configurations. Currently this is not possible in AMPT simulation which does not contain spin information and produces them all together. In order to  separate these channels, we    estimate  the ratio of yields between two such channels, e.g. $A$ and $B$ with mass $M_{A}$ and $M_B$ (either color neutral or colored ones). A reasonable method is to use  thermal model relation:
 \begin{eqnarray}
 R\equiv \frac{\rm Yield(A)}{\rm Yield(B)} = \exp\left(\frac{M_{B}-M_{A}}{T}\right),
 \label{eq:R}
 \end{eqnarray}
with a temperature parameter $T=160~\mathrm{MeV}$~\cite{Andronic:2019wva}. 
For the hadronic molecule picture, $A$ and $B$ are the $D^*$ and the $D$ mesons, respectively~\cite{Tanabashi:2018oca}. 
For tetraquark picture, they are for the spin triplet $[cq]_1$ diquarks and the spin singlet $[cq]_0$ diquarks
{(The diquark masses, taken from Refs.~\cite{Anisovich:2010wx,Anisovich:2015caa}, are extracted by  fitting to known baryons' mass spectrum. Notice that in tetraquark picture~\cite{Maiani:2014aja}, the mass values of the higher $0^{++}$, the higher $1^{+-}$ and the $2^{++}$
 tetraquarks coincide with each other. In this study, the $0^{++}$ is presented as an illustration.)}, respectively. 
 This estimate indicates a composition of ($30\% , 70\%$) for ($D^*$, $D$) and a composition of  ($35\% , 65\%)$ for spin (triplet, singlet) diquarks, which will be used in our simulations. 
 Such composition   depends on the parameter $T$ in the above equation, for example, varying $T$ from 130 to $190~\mathrm{MeV}$, the ($D^*$, $D$) composition evolves from ($25\% , 75\%)$  to ($32\% , 68\%)$.  
  To quantify the impact of this composition, we will   show uncertainty bands by comparing results from varying this composition up and down by $10\%$. 

We note that our calculations only account for late-time production of $\x$  at freeze-out time and neglect potential contributions from $c\bar{c}$ pairs that are produced early and that survive through the plasma. 
Efforts are being made to overcome this limitation of the AMPT framework and allow quantitative study of initial production as well.
Such contribution would be negligible for molecular $\x$  that have large size and could be easily dissolved by hot medium.  It may, however, be an important addition to the yield in the case of tetraquark $\x$. Based on quantitative simulations of the initial production versus regeneration for  $J/\Psi$ at LHC~\cite{Zhou:2014kka,Zhao:2020jqu,Zhao:2011cv,Du:2015wha,Rapp:2008tf}, it would be reasonable to expect that adding such contribution would at most double the yield of tetraquark $\x$ production from our results.

{\it Results.---} 
With this simulation framework, we have generated a total of one million minimum  bias events for Pb-Pb collisions at $\sqrt{s} = 2.76$ TeV. The inclusive yield of $\x$ is computed to be  around $220\,000$ in the molecule scenario while to be around $900$ in the tetraquark scenario. A pronounced finding is significantly more production  of   the molecule state than that of the tetraquark state, by a factor of $250$ --- a 2-order-of-magnitude difference. This result may be understood as follows: $c$ and $\bar{c}$ quarks are carried by bulk flow, randomly diffuse around the whole fireball volume, and, in general, would be somewhat separated in space by the time of freeze-out;  in the molecular picture, the constituents $D^*$ ($\bar{D}^*$) and  $\bar{D}$ ($D$) (containing either a $c$ or $\bar{c}$ quark) prefer to form $\x$ when they are well separated; in the tetraquark picture, the constituents diquark and antidiquark (each also containing a $c$/$\bar{c}$ quark) needs to stay very close in space; 
as such, there is a much higher probability for the formation of hadron molecules than tetraquark states.  
We note this is different from the production of charmonium states where the comparison, e.g., between ground state $J/\Psi$ and excited state $\Psi(2s)$ is dictated by their different binding energy.

\bef
\psfig{file=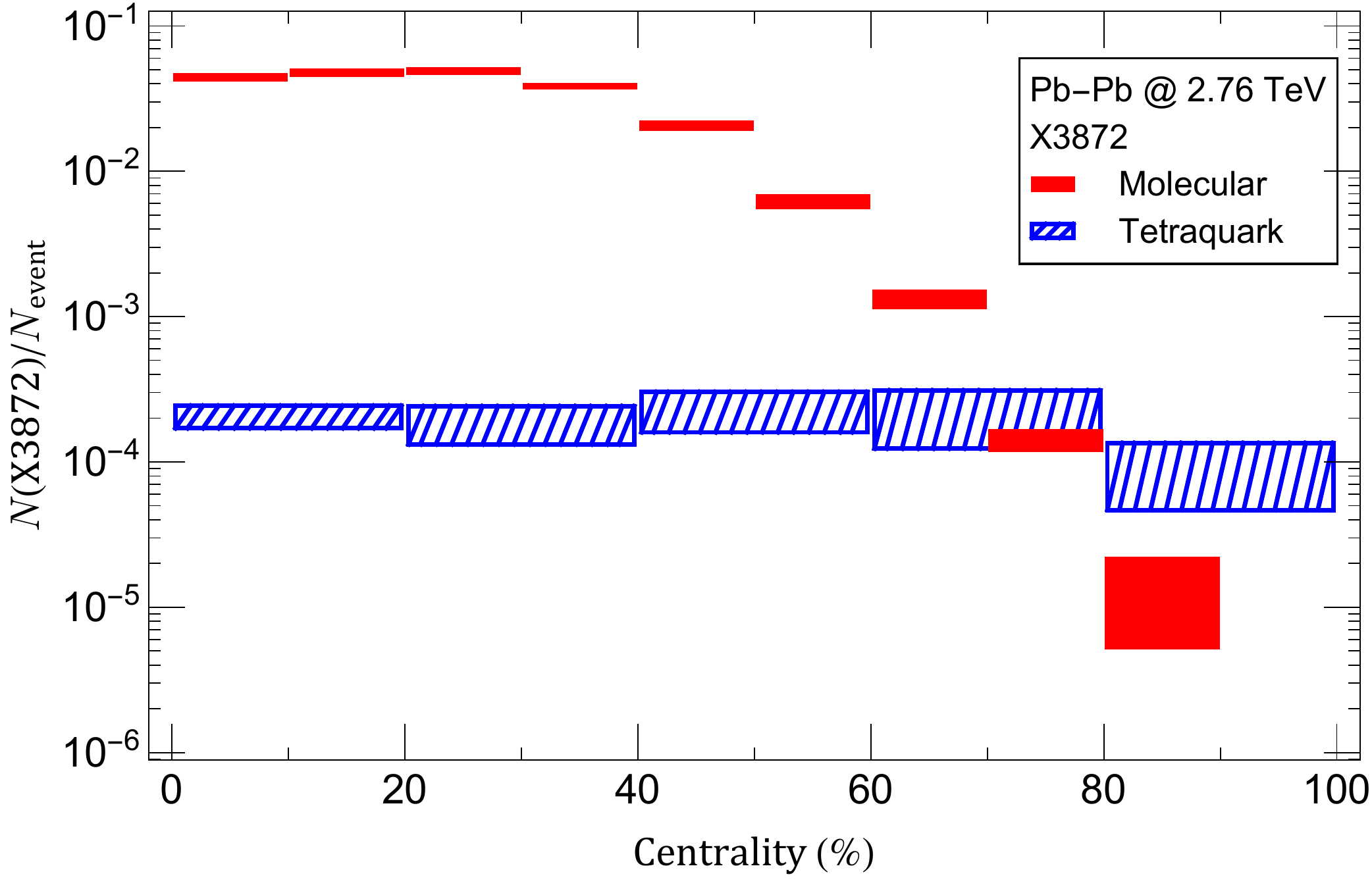, width=3.3in}
\caption{The centrality dependence of the $\x$ in Pb-Pb collisions at $\sqrt{s}=2.76~\mathrm{TeV}$ for hadronic molecular configuration (red solid boxes) and tetraquark configuration (blue shaded boxes), computed from our framework. The bands reflect both statistical uncertainty from our simulations and the uncertainty due to constituent composition as discussed around Eq.~\eqref{eq:R} that are obtained from varying the composition fraction by $\pm 10\%$.   }
\label{fig-cent}
\eef 

This interpretation appears to be further confirmed by the centrality dependence of the $\x$ yield shown in Fig. \ref{fig-cent}. Going from central to peripheral collisions, one observes a strong decrease  for the molecular scenario while a  mild change for the tetraquark scenario. As a baseline of expectation, the available number of $c$ and $\bar{c}$ quarks would gradually decrease with increasing centrality class, with the fireball spatial volume and evolution time also decreasing. The sharp decrease of molecular state production toward very peripheral collision is due to the shrinking volume available for accommodating the large-size hadronic molecule.  The relatively flat dependence of the tetraquark case is due to two compensating factors: decreasing numbers of $c/\bar{c}$ quarks while increasing chances of small spatial separation between (anti-)diquarks due to shrinking fireball volume. Such observation suggests that it would be a good idea to probe the system-size dependence of $\x$ production, e.g., by measuring them across colliding systems like Pb-Pb, Au-Au, Xe-Xe, Cu-Cu, O-O, $d$-$A$/$p$-$A$, etc.

\bef
\psfig{file=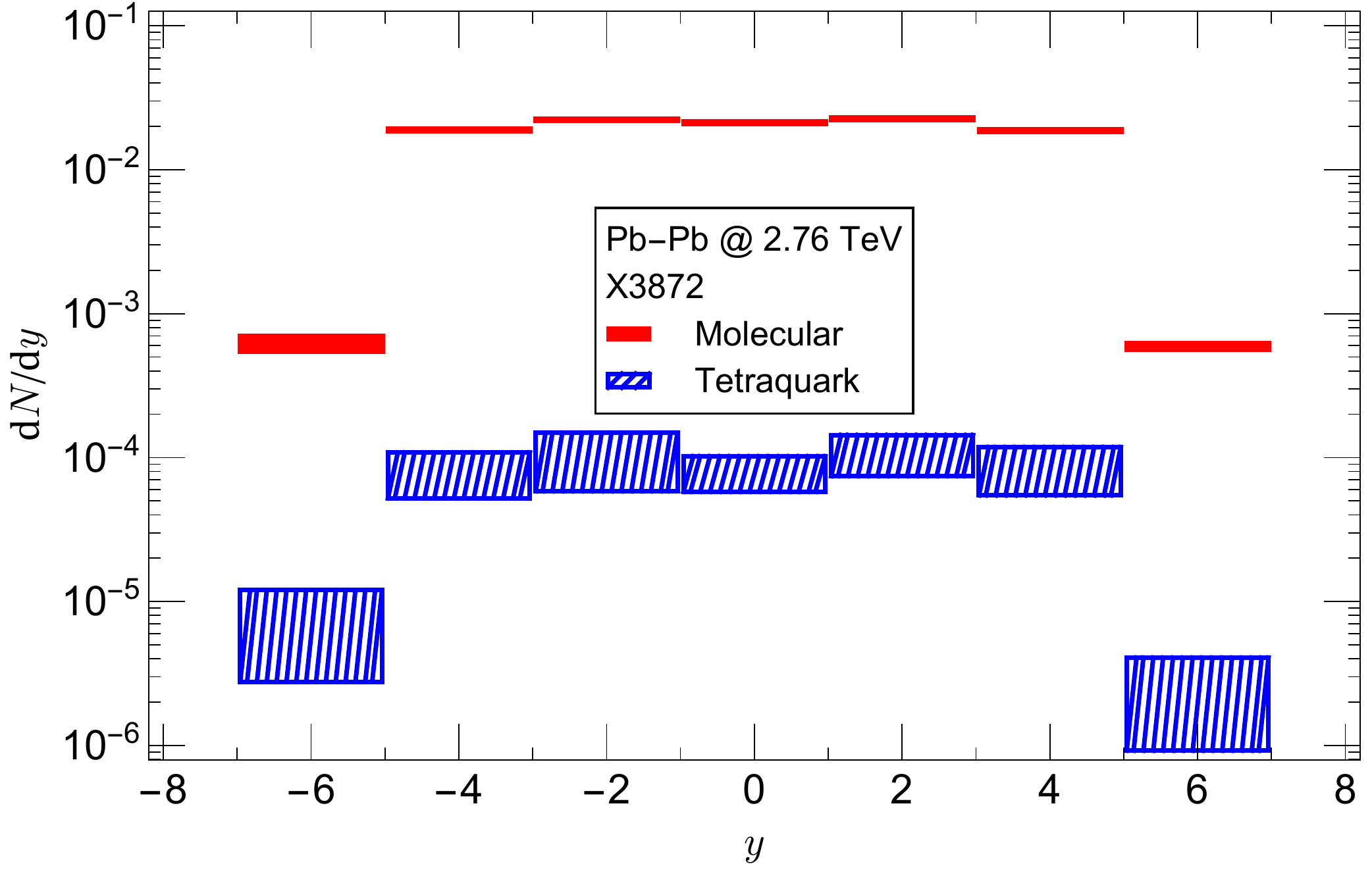, width=3.3in}
\caption{Rapidity distribution of the $\x$ yield in Pb-Pb collisions at $\sqrt{s}=2.76~\mathrm{TeV}$ for hadronic molecular configuration (red solid boxes) and tetraquark configuration (blue shaded boxes), computed from our framework. The bands are similarly determined as described in Fig.~\ref{fig-cent}.}
\label{fig-rapidity}
\eef

\bef
\psfig{file=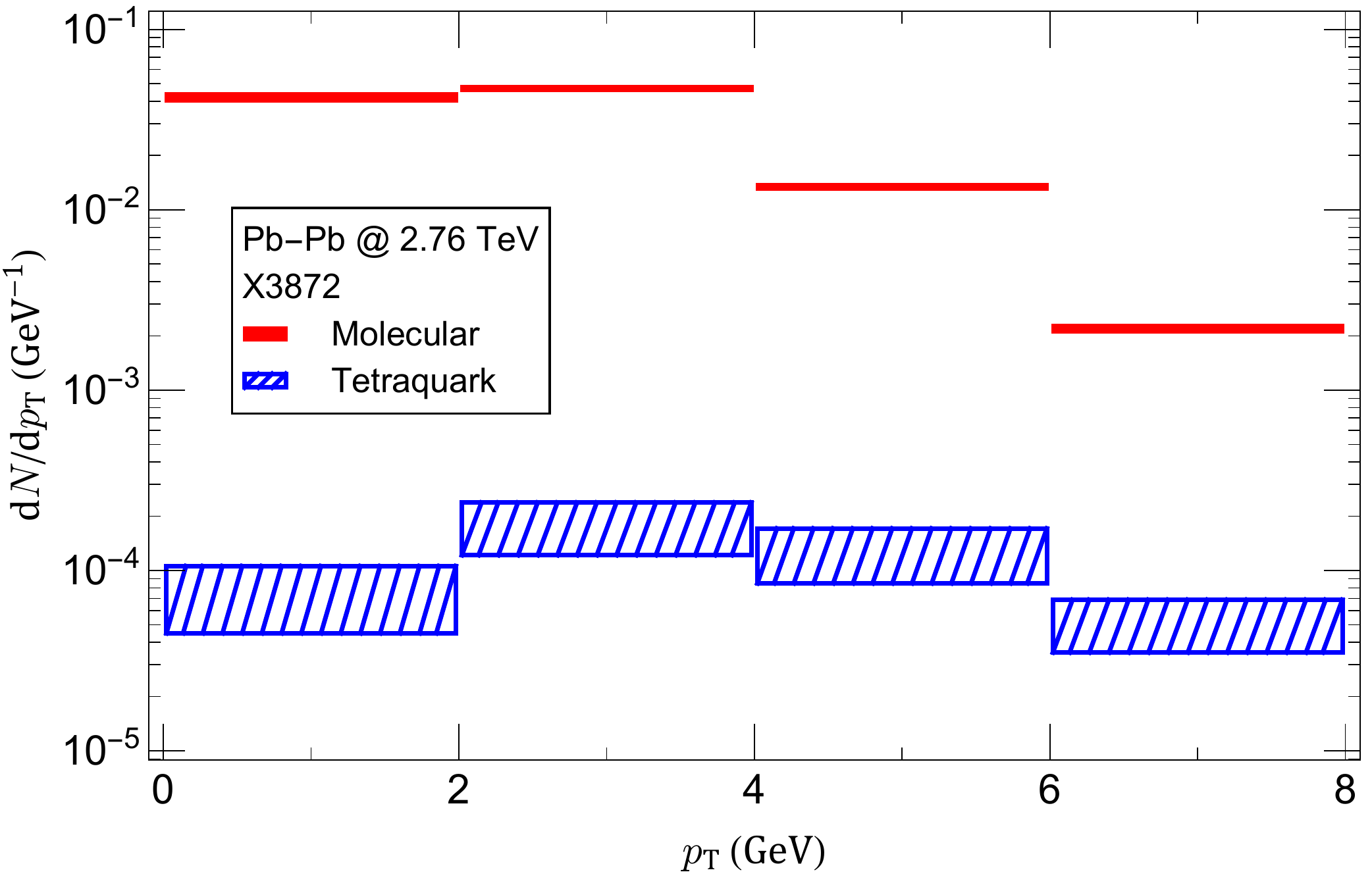, width=3.3in}
\caption{Transverse momentum spectra of the $\x$ yield in Pb-Pb collisions at $\sqrt{s}=2.76~\mathrm{TeV}$ for hadronic molecular configuration (red solid boxes) and tetraquark configuration (blue shaded boxes), computed from our framework. The bands are similarly determined as described in Fig.~\ref{fig-cent}.}
\label{fig-pt}
\eef

We next present the rapidity distribution of the $\x$ production in Fig.~\ref{fig-rapidity} as well as the transverse momentum spectra in Fig. \ref{fig-pt} in minimum bias Pb-Pb collisions. The rapidity dependence of both scenarios is similar to that of various normal hadrons~\cite{Chatrchyan:2011pb,Abbas:2013bpa}, being relatively flat in the region close to central rapidity  while decreasing toward more forward/backward region. Even though   the molecular $\x$ has a large size, the two coalescing constituent $D$ mesons are typically close in rapidity as otherwise they would have a large relative momentum and could not easily satisfy the mass constraint.
The  $p_\mathrm{T}$ spectra of the $\x$  show a similar overall trend to those for normal hadrons and are indicative of production from thermal source with radial flow. The tetraquark case shows a harder slope at higher $p_T$ than the molecular case. The reason could be   that   the diquark and anti-diquark in a tetraquark-$\x$ are from close-by fluid cells with more collimated  flow and can more easily add together to form a larger-$p_T$   $\x$  particle.

\bef
\psfig{file=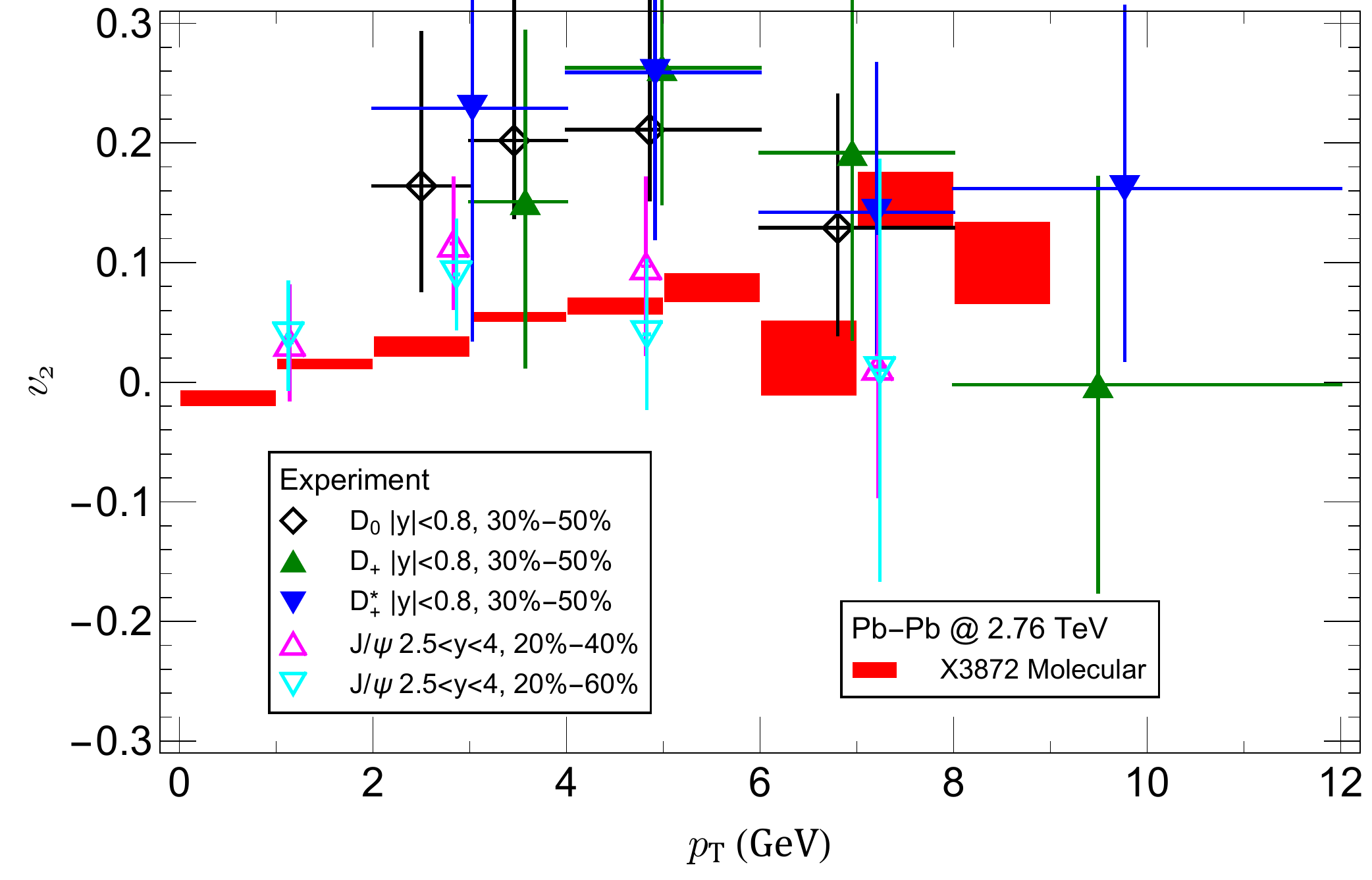, width=3.3in}
\caption{The elliptic flow coefficient $v_2$ versus transverse momentum $p_T$ for produced $\x$  in minimum bias Pb-Pb collisions at $\sqrt{s}=2.76$ TeV, predicted from our computation for the hadronic molecule picture. The bands are similarly determined as described in Fig.~\ref{fig-cent}. These results are compared with experimental data for $D$ mesons and $J/\Psi$ elliptic flow at the same collision energy.}
\label{fig-v2}
\eef 

One interesting  question is are the produced $\x$ hadrons part of the collective flow?  To this end  the anisotropic flows would be the key observables. The first such result, for $\x$ elliptic flow $v_2(p_T)$, is shown in Fig. \ref{fig-v2} and compared with experimental data for $v_2$ of $J/\Psi$ and $D$ mesons~\cite{Sirunyan:2018toe,Acharya:2017tgv, Abelev:2013lca, ALICE:2013xna}. 
Within 1M event, the limited statistics 
would only allow a meaningful evaluation for the  molecule case. 
Our result predicts a considerable elliptic flow for the produced $\x$ with a characteristic $p_T$ dependence  similar to other hadrons. 
We compare the result with measured $v_2$ of  $J/\Psi$, which also contains $c$/$\bar{c}$ and has a mass value not far from $\x$. The computed $v_2$ of $\x$  is comparable to  that of $J/\Psi$ (within the very large error bars). Our next comparison is with $D$ mesons. The molecule state is formed via coalescing two $D$ mesons. If the $\x$ size were to be compact, these two constituents would be from nearby fluid cells with their flow effect added coherently into the $\x$, in a way similar to the well-known constituent quark scaling observed in light and strange hadron elliptic flow~\cite{Sirunyan:2018toe}. Instead the constituent scaling would break down for $\x$ if it has a large size with two $D$ mesons originating from remote patches of the fluid. Our computed $\x$ elliptic flow is smaller than the $D$ meson $v_2$ data, in consistency with a large size hadron molecule. Future measurement of the $\x$  elliptic flow   would be highly interesting to help decipher its nature.


 {\it Summary.---}
In this work, we have demonstrated the novel opportunity to explore the nature of the $\x$ in heavy ion collisions. Through implementing production mechanism for $\x$ either as hadronic molecule or as compact tetraquark on top of  bulk medium evolution, we have made quantitative predictions in both scenarios for a series of  heavy ion observables 
which will provide valuable guidance for experimental programs. We particularly propose to measure the elliptic flow of the $\x$, which is computed for the first time and found to be sizable. A major highlight of our results is that  the fireball volume is a key factor in the production of $\x$, leading to about 2 orders of magnitude higher yield as well as a significantly stronger centrality dependence when assuming its structure to be a hadronic molecule than that for a compact tetraquark. Such tantalizing findings could potentially open a new path for deciphering the nature of $\x$ via heavy ion measurements.  

All these results together provide a multitude of predictions characterizing the $\x$ production in heavy ion collisions, which shall strongly motivate enthusiastic experimental activities in the near future. Recent CMS measurements show an interesting enhancement of the $\x$-to-$\Psi(2S)$ ratio for $15 < p_T<50$ GeV in Pb-Pb collisions comparing to proton-proton collisions \cite{CMS:2019vma}. Such data are highly suggestive of the potential medium effect on $\x$ production in the high $p_T$ region, while not suitable for comparison with our  simulation results that focus on the soft bulk production. Efforts are underway to extend our study toward the high $p_T$ region and one would also expect future experimental data for $\x$ production in the soft region. 
The present exploratory study shall also lead to further theoretical investigations such as calculating the production of other exotic candidates (e.g., pentaquarks) in heavy ion collisions, improving the formation mechanism by including spin degrees of freedom, evaluations of these states within hydrodynamic  model, etc. 
 It is tempting to envision an exciting time of vibrant and coherent theory and experiment efforts for exploring heavy ion collisions as a massive production factory of exotic hadrons to its fullest extent. 

\medskip

\begin{acknowledgements}
The authors thank P.~Braun-Munzinger, S.~Cho, Y.J.~Lee, Z.-W.~Lin, M.~Shepherd, A.~Szczepaniak, and L.~Zheng for useful discussions and communications.
This work is supported in part by Guangdong Major Project of Basic and Applied Basic Research No.~2020B0301030008 and by the National Natural Science Foundation of China (NSFC) under Grants No. 12022512 and No. 12035007, and by the Deutsche Forschungsgemeinschaft (DFG) through the funds provided to the Sino-German Collaborative Research Center ``Symmetries and the Emergence of Structure in QCD"  (NSFC Grant No. 11621131001 and DFG Grant No. TRR110),
 by the NSFC Grants No. 11735007 and by the NSF Grant No. PHY-1913729. 
This work is also supported by the Science and Technology Program of Guangzhou under Contract No. 2019050001 and Guangdong Provincial funding with No. 2019QN01X172.

\end{acknowledgements}


\end{document}